\documentclass[12pt,aps,nofootinbib]{revtex4}
\usepackage[utf8]{inputenc}
\usepackage{amsmath}
\usepackage{amsfonts}
\usepackage{amssymb}
\usepackage{graphicx}
\usepackage{grffile}
\usepackage{color}
\usepackage{textgreek}
\usepackage{hyperref}
\usepackage{multirow}
\usepackage[normalem]{ulem}
\usepackage{natbib}
\begin{document}
\title{Collective dynamic length increases monotonically in pinned and unpinned glass forming systems}
\author{Rajsekhar Das$^{1}$}
\author{T. R. Kirkpatrick$^{2}$}
\author{D. Thirumalai$^{1}$}
\affiliation{
$^1$ Department of Chemistry, University of Texas at Austin, Austin, Texas 78712, USA
}
\affiliation{
$^2$ Institute for Physical Science and Technology, The University of Maryland, College Park, MD 20742, USA
}

\begin{abstract}
The Random First Order Transition Theory (RFOT)  predicts that transport proceeds by cooperative movement of particles in domains whose sizes increase as a liquid is compressed above a characteristic volume fraction, $\phi_d$. The rounded dynamical transition around $\phi_d$, which signals a crossover to activated transport, is accompanied by a growing correlation length that is predicted to diverge at the thermodynamic glass transition density ($> \phi_d$). Simulations and imaging experiments probed the single particle dynamics of mobile particles in response to pinning all the particles in a semi-infinite space or randomly pinning (RP) a fraction of particles in a liquid at equilibrium. The extracted dynamic length increases non-monotonically with a peak around $\phi_d$, \textcolor{black}{which not only depends on the pinning method but is different from $\phi_d$ of the actual liquid}. This finding is at variance with the results obtained using the small wave length limit of a four-point structure factor for unpinned systems. To obtain a consistent picture of the growth of the dynamic length, one that is impervious to the use of RP, we introduce a multi particle structure factor, $S^c_{mp}(q,t)$, that probes collective dynamics. The collective dynamical length, calculated from the small wave vector limit of $S^c_{mp}(q,t)$, \textit{increases monotonically} as a function of the volume fraction in glass forming binary mixture of charged colloidal particles in both unpinned and pinned systems. This prediction, which also holds in the presence of added monovalent salt,   may be validated using imaging experiments.   
\end{abstract}
\maketitle

\vskip 1 cm
\section{Introduction}
In a landmark article published in 1964, Aneesur Rahman \cite{Rahman64PhysRev} reported the results of molecular dynamics (MD) simulations investigating the collective motions of atoms in liquid Argon. The simulations, performed using CDC (Control Data Corporation) 3600 computer, consisted of 864 particles at 94.4$^{\circ}$K  and at the density 1.374 $g \cdot cm^{-3}$. For reference, the temperature and density at the triple point of Argon are 80.8$^{\circ}$K and 1.417 $g \cdot cm^{-3}$, respectively.  In retrospect, it is clear that the 1964 study laid the foundation for using MD simulations in ways that were unimaginable at that time. Rahman also developed the theoretical basis of MD simulations and applied them to a bewildering range of problems, spanning phase transitions in solids~\cite{Parrinello1981JAP,Ray1985JCP}, lattice gauge simulations \cite{Callaway82PRL}, structure of molecular fluids, especially liquid water \cite{Rahman71JCP,Stillinger74JCP},  aspects of glasses \cite{Rahman76JCP,Nagel81PRL}, and much more. His pioneering work with Rossky and Karplus showed that MD \cite{Rossky79Biopolymers} could be used to simulate peptides and proteins in aqueous solutions, which have, over the last nearly five decades,  provided quantitative insights into their dynamics that are often difficult to obtain from experiments alone. These and other studies by Rahman, which are far too many to enumerate here, propelled  MD as an indispensable branch of study in physical sciences and biology.

Not surprisingly, molecular dynamics simulations have been instrumental in testing different theories of the structural glass transition (SGT)~\cite{AdamGibbs1965,KirkpatrickPRA1989,ThirumalaiRMP2015,Tarjus_2005,BiroliJCP,Lubchenko07ARPC}.  Given that current  the literature is vast, we cite only a few references including some recent reviews  \cite{Bernu87PRA,Mountain87PRA,Barrat91ARPC,Kob95PRE,Scalliet22PRX,Barrat23ComRendus}. 
Among the contenders, the Random first-order transition theory (RFOT)~\cite{Kirkpatrick2023,KirkpatrickPRA1989,Lubchenko07ARPC,Bouchaud2009,ThirumalaiRMP2015} has been successful in explaining  most of the phenomena associated with SGT. (1) RFOT predicts that there ought to be two transitions as the system is cooled by lowering the temperature, $T$ or compressed by increasing the density. Upon supercooling (or compressing) a liquid, a rounded dynamical transition occurs at the temperature $T \approx T_d$ (or $\phi_d$) ($T_d$ ($\phi_d$) is a roughly the mode-coupling transition temperature (density)).  
 The liquid is pictured as a ``mosaic'' of correlated amorphous domains at volume fractions exceeding $\phi_d$ . 
 Above $\phi_d$ transport occurs by activated processes, although typically barrier limited transport takes place even before $\phi_d$ is reached because of the rounded nature of the transition at $\phi_d$. The driving force for transport, above $\phi_d$, is entropic in origin and is opposed by scale dependent surface tension~\cite{KirkpatrickPRA1989}. The latter is a unique feature of RFOT.   
At a lower temperature or higher density a thermodynamic ideal glass transition occurs. 
(2) Due to the emergence of multiple mosaic domains, the dynamics, which become heterogeneous around $\phi_d$, are accompanied by a growth in the dynamic correlation length.  RFOT predicts both the static and dynamic correlation lengths should diverge at the ideal glass transition. 

Estimation of the dynamic  or static correlation length  using  MD simulations is challenging, even for $\phi \sim \phi_d$ ($\phi$ is the volume fraction in the colloidal particles) because of the difficulty in equilibrating the sample within the simulation time. The use of the particle pinning technique (an uncontrolled externally imposed disorder that breaks translational invariance )~\cite{RP1, RP2, RP3, RP4, BiroliJCP, Cammarota8850, Das2016, Ozawa6914}, makes it possible to obtain equilibrium data close to or even above $\phi_d$.  However, estimates of the dynamic length close to $\phi_d$ (or below $T_d$) using the RP method using simulations and experiments have led to contradictory results.  
By pinning all the particles in a binary mixture of harmonic spheres in a semi-infinite space the response in the other half-space was probed~\cite{Kob2012NatPhys}.  The use of a frozen amorphous wall is a generalization of the point-to-set method~\cite{Biroli04JCP,Biroli08NatPhys,Hocky12PRL} for estimating the dynamic length. The calculated   length ({$\xi_d^s$}) using single particle  dynamics  changed  \textit{non-monotonically} with a peak at $\approx T_d$ as $T$ is decreased.

Flenner and Szamel (FS)~\cite{Flenner12NatPhys} simulated the same model without RP and calculated $\xi_d$ by fitting the low wave vector behavior of the four-point dynamic structure factor $S_4^s(q,t)$ to  the Ornstein-Zernike equation (OZ). In contrast to the previous simulations~\cite{Kob2012NatPhys}, they found that $\xi_d^s$ increases  \textit{monotonically}. As pointed out by FS, the use of RP, especially the amorphous wall, elicits (unknown) non-linear response on the dynamics of the \textcolor{black}{unpinned} particles. For this reason, it is unclear if the RP method faithfully reports the dynamics of unperturbed bulk liquid~\cite{Flenner12NatPhys}. 

Subsequent simulations using a binary mixture of particles interacting via harmonic ~\cite{MeiPRE2017}  and Lennard-Jones (LJ) potentials ~\cite{HockyPRE2014} have added to the discussion on the usefulness of the RP method. By placing two walls separated by a distance, $d$, it was shown that the temperature at which $\xi_d$ attains a maximum is $d$ dependent~\cite{MeiPRE2017}. More strikingly, the maximum vanishes if the walls are smooth. Simulations of the LJ system revealed that $\xi_d$, determined by using the point-to-set method~\cite{Biroli08NatPhys},  saturates when $T > T_d$ without showing the expected non-monotonic behavior.

Two insightful experiments that support the thermodynamic origin of the SGT appeared after the simulations~\cite{Kob2012NatPhys,Flenner12NatPhys}. (1) An experimental realization of a frozen wall of particles was achieved by a holographic optical tweezer setup~\cite{HimaNagamanasa15NatPhys}. Video microscopy experiments on a binary mixture of micron-sized neutral polystyrene particles in two dimensions were used to measure the response of the frozen amorphous wall on the dynamics as a function of the distance from the wall.  By fitting the $z$ (distance from the wall) dependence of the time-dependent decay of the self part of the overlap function, it was found that the inferred dynamic length increases non-monotonically as a function of the packing fraction, which was in accord with the simulation results~\cite{Kob2012NatPhys}, obtained by changing the temperature. 
(2) Because randomness is self-generated in glasses~\cite{Kirkpatrick89JPhysA}, it would prudent to realize pinned configurations by probing the response of motile particles to ``slow" particles that relax on times that are much greater than the typical relaxation time, $\tau_\alpha$.   Such an experiment was performed on a binary mixture of colloidal polystyrene particles~\cite{Ganapathi18NatComm} in three dimensions. They first identified particles that mimic pinned particles (self-induced pinning) for which the relaxation time is on the order of $(5-7) \tau_\alpha$. The extracted $\xi_d$ also changes non-monotonically, with a maximum value $\xi_d \approx 10 \sigma_s$ ($\sigma_s = 1.05 \mu m$ is the size of the smaller particle) as the volume fraction increases.  It is worth pointing out that in both these experiments there is only one data point in the $\xi_d$ versus packing fraction plots after the maximum near $\phi_d$ (Fig. 2c in \cite{Ganapathi18NatComm} and Fig. 3a in \cite{HimaNagamanasa15NatPhys}). The results were interpreted in terms of changes in shapes of the mosaic domains \cite{Stevenson06NatPhys} as $\phi$ approaches $\phi_d$. \textcolor{black}{Despite changing the pinning protocol the qualitative conclusions in the three studies \cite{Kob2012NatPhys,HimaNagamanasa15NatPhys,Ganapathi18NatComm} are similar - the dynamical length increases non-monotonically reaching a maximum at a temperature or density that is close to the dynamic transition point of the pinned system.}


The foregoing summary of the experimental and simulation results shows that the extraction of  $\xi_d$ using the RP method and its interpretation is not straight forward.  To obtain consistent results for the dynamic length, without and with RP, we introduce a two-point overlap function, which has been used to distinguish between the folded and unfolded states of globular proteins~\cite{Camacho93PNAS,guo1995kinetics}. Using simulations of binary mixtures of charged colloidal suspensions~\cite{Hyun2019}, which form classical Wigner glasses~\cite{lindsay82JCP},  we calculated the collective multi-point dynamic structure factor $S^c_{mp}(q,t)$, describing the fluctuations in the two-point overlap function, as well as the self-part, $S^s_4(q,t)$.   The main findings may be summarized as follows.
    (1) Collective  length  (referred to as $\xi_d^c$ here on) obtained by fitting the low $q$ part of $S^c_{mp}(q,t)$ to the OZ equation grows \textit{monotonically} as a function of $\phi$ in both pinned and unpinned  systems. 
In contrast, $\xi_d^s$ calculated using the low $q$ dependence of the self part of four-point dynamic structure factor $S^s_4(q,t)$  grows \textit{non-monotonically} with a peak near $\phi_d$. 
(2) Strikingly, for the unpinned liquid, the length scale calculated from both $S^s_4(q,t)$ and $S^c_4(q,t)$ gives quantitatively identical values,  exhibiting monotonic growth as a function of $\phi$. On the other hand, in a pinned system, $\xi_d^s$ ($\xi_d^c$) increases non-monotonically (monotonically). 
(3) For the pinned liquids we show that the collective dynamics become increasingly important as $\phi$ increases. Single particle dynamics, used in all previous studies \cite{Kob2012NatPhys,HimaNagamanasa15NatPhys},  underestimates correlated dynamics, which is especially relevant as $\phi$ approaches $\phi_d$.
(4) At the onset of collective dynamics, which occurs at $\phi < \phi_d$  there is a dynamic crossover that can be probed using RP. The relaxation time $\tau_\alpha$ as a function of $\xi^c_d$ changes from $\tau_\alpha \sim \exp[k\xi^c_d]$ to $\tau_\alpha \sim \exp[k(\xi^c_d)^{3/2}]$~\cite{Flenner12NatPhys} in accord with the predictions of RFOT~\cite{KirkpatrickPRA1989}. The $(\xi^c_d)^{3/2}$ scaling follows from very general arguments \cite{Kirkpatrick2023} and is independent of the method used to compute the length as function of $\phi$.

\section{Models and Methods}
\textbf{Charged colloids:} We simulated a binary mixture of highly charged micrometer-sized colloidal particles~\cite{Kang13PRE,Rosenberg_1989,Hyun2019}, which form classical Wigner glasses \cite{lindsay82JCP}, by  randomly pinning  a fraction ($\rho_{\text{pin}}$)  of particles at their equilibrium positions. We  also simulated unpinned ($\rho_{\text{pin}}=0$) systems. The 50:50 binary mixture is composed of two types of particles with diameters $a_1 = 5.5\mu m$ and $a_2 = 1.1\mu m$. 

The  particles, separated by a distance $r_{ij}$, interact via the Derjaguin-Landau-Verwey-Overbeek potential~\cite{DLVO1,DLVO2,DLVO3,DLVO4,DLVO5},
\begin{equation}
V_{ij}(r_{ij}) = \frac{e^2Z_iZ_j}{4\pi\epsilon}\left(\frac{\exp[\kappa a_i]}{1+\kappa a_i}\right)\left(\frac{\exp[\kappa a_j]}{1+\kappa a_j}\right)\left(\frac{\exp[-\kappa r_{ij}]}{r_{ij}}\right),
\label{DLVO}
\end{equation}
where $Z_i$, $\kappa$ and $\epsilon$ are the valence of the charged colloids, inverse Debye-H\"{u}ckel screening length, and the dielectric constants, respectively. The values of $Z_i$ for the large and small particles are $600$ and $300$, respectively. The dielectric constant $\epsilon = \epsilon_0\epsilon_r$, where $\epsilon_0$ is the free space permittivity, and $\epsilon_r$ is the relative permittivity. At the temperature $T=298 K$, we  set the dielectric constant to $\epsilon_r = 78$. The inverse Debye-H\"{u}ckel screening length $\kappa$ is,
\begin{equation}
\kappa^2 = \frac{e^2}{\epsilon K_B T}\left(\rho_c z_c^2 +\sum_{i'}^{n}\rho_{i'}z_{i'}^2\right),
\label{kappa}
\end{equation}
where $\rho_c$ and $z_c$ are the number density and valence of the counterions, respectively,  $\rho_i$ and $z_i$ are the number density and the valence of the added salt, respectively, and $k_B$ is the Boltzmann constant. For monovalent ions, $|z_c| =1$ and $\rho_c = \rho_1 Z_1 + \rho_2 Z_2$, where $\rho_1$ and $\rho_2$ are the number density of the small and the large particles respectively. In the presence of added monovalent ions, $\rho_{add} = \sum_{i'}^{n = 2}\rho_{i'}$,   we define a relative excess ion density, 
\begin{equation}
    \rho_r = \tfrac{\rho_{add}}{\rho_c}.
    \label{rhor}
\end{equation}
In the current work, we set $\rho_r =0$ and  $5$. The effective range of interaction increases with decreasing $\phi$ and $\rho_r$~\cite{Hyun2019}. The interaction potential $V_{ij}(r)$ is truncated at a  distance $r_c$, determined by the condition $V_{ij}(r_c) = 0.001k_BT$. Because the interaction potential is not as long-ranged as the Coulomb potential, we did not find the need to use Ewald summation.  
The thermodynamic variable in the simulations is the volume fraction,
\begin{equation}
\phi =\frac{4\pi}{3L^3}\left(N_1a_1^3 + N_2a_2^3\right).
\label{pf}
\end{equation}
Total number of particles, $N = N_1+N_2$, is typically 10,000.

\subsection*{Methods}
\noindent We  performed Brownian dynamics ($BD$) simulations by integrating the  equation of motion,
\begin{equation}
\frac{d\vec{r}_i(t)}{dt} = -\nabla_{r_i}U(\vec{r}_1,...,\vec{r}_N)\frac{D_{i,0}}{K_BT}\delta t +\sqrt{2D_{i,0}}\vec{R_i}(t),
\label{equim}
\end{equation}
where, $\vec{r_i}(t)$ is the position of the $i^{th}$ particle at  time $t$, $U(\vec{r}_1,...,\vec{r}_N) = \sum_{i\neq j}V_{ij}(r_{ij})$. The integration time step is $\delta t$, and $D_{i,0}$ is the bare diffusion constant of the $i^{th}$ particle. The bare diffusion constants for the small and the large colloidal particles are chosen as $4.53 \mu m^2/s$ and $2.24 \mu m^2/s$, respectively. We calculated $D_{i,0}$  using the Stokes-Einstein relation, $D_{i,0} = \frac{k_BT}{6\pi\eta a_i}$, where $\eta = 0.89$ mPa s for water.  The $\vec{R}_i(t)$ on the right-hand side of the Eqn.~\eqref{equim} is a random noise, which satisfies $\left\langle \vec{R}_i(t)\vec{R}_j(t')\right\rangle = 6D_{i,0}\delta_{ij}\delta(t-t')$, where $\delta_{ij}$ is the Kronecker delta and $\delta(t-t')$ is the Dirac delta function. The integration time step $\delta t$ is set to $10$ $\mu s$ for $\rho_r =0$ and $1$ $\mu s$ for $\rho_r = 5$ to ensure numerical accuracy~\cite{Hyun2019}. The relative excess ion density $\rho_r = 5$ corresponds to $\approx 0.44$ millimolar salt concentration.

We reported simulation results for these systems in a previous study~\cite{Hyun2019} without RP. The calculated values of  $\phi_d$ for $\rho_r = 0$ is $\approx 0.10$ and for $\rho_r = 5$ is $\approx 0.36$. The increase in $\phi_d$ reflects a decrease in the range of DLVO potential.


\section{Results \& Discussions}
\begin{figure}[ht!]
\begin{center}
\includegraphics[width=0.98\columnwidth]{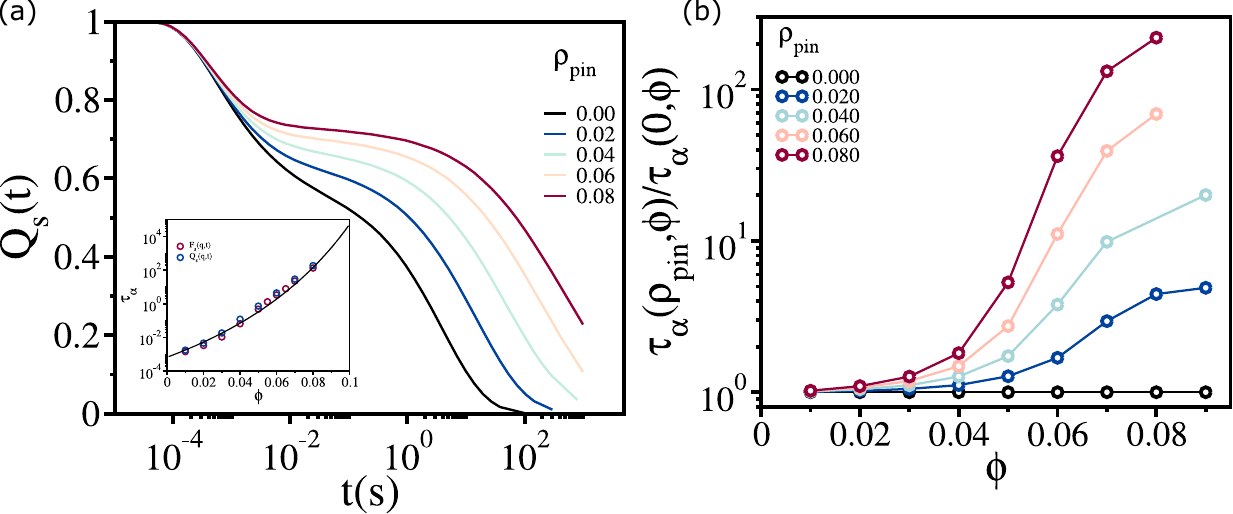}
\caption{ {\bf  Structural relaxation time in randomly pinned systems:} (a) Self overlap function $Q_s(t)$ as function of $t$ for $0.00\leq \rho_{\text{pin}}\leq0.08$ at $\phi = 0.09$ for $\rho_r =0$. Inset compares $\tau_\alpha$ calculated using $Q_s(t)$ and the self-intermediate scattering function, $F_s(q,t)$, with $q = 2\pi/d_{\text{ave}}$. (b) $\tau_\alpha(\rho_{pin},\phi)/\tau_\alpha(0,\phi)$ as a function of $\phi$ for $0.00\leq\rho_{\text{pin}} \leq 0.08$ and $\rho_r = 0$. }
\label{ratio}
\end{center}
\end{figure}
\noindent \textbf{Pining enhances the relaxation times:} We calculated the relaxation time, $\tau_\alpha$, using the decay of the self structural overlap function,
\begin{equation}
    Q_s(t)=\frac{1}{N-N_p}\left[\left\langle\sum_{i=1}^{N-N_p}w_i(t)\right\rangle\right],
    \label{Qst}
\end{equation}
where $\left[\left\langle ...\right\rangle\right]$ is the ensemble average and average over different time origins, $N$ is the total number of particles, and $N_p$ is the total number of pinned particles. The overlap function,  $w_i(t) = \Theta\left(a -|\vec{r}_i(t) -\vec{r}_i(0)| \right)$, describes the duration for which particle $i$ remains within a distance $a$ at time $t$ given that it is within $a$ at $t =0$; $\vec{r}_i(t)$ is the position of a particle at time $t$. The Heaviside function, $\Theta(x)$, is unity if $|\vec{r}_i(t) -\vec{r}_i(0)| \leq a$ and zero otherwise. Thus, $Q_s(t)$ measures the average number of particles that do not move a distance of $a$ from their original position after a time $t$. These particles could be referred to as slow particles. Note that, $Q_s(t)$ has similar information as the self-intermediate scattering function $F_s(q,t) = N^{-1}\left\langle\sum_{i= 1}^N e^{-q\cdot[\vec{r_i}(t) - \vec{r_i}(0)]} \right\rangle$~\cite{OZ1}. With the choice of $a = 0.3d_{\text{ave}}$~\cite{Hyun2019}, we find that $\left\langle Q_s(t)\right\rangle$ decays with a relaxation time $\tau_\alpha$, which coincides with the time scale of decay of the self intermediate scattering function $F_s(q,t)$ with $q = 2\pi/d_{\text{ave}}$ (Fig.~\ref{ratio} (a) inset). The volume-averaged value of the diameter of the charged colloids is $d_{\text{ave}} = 2\frac{\phi_1a_1 + \phi_2a_2}{\phi_1+\phi_2}$ with $\phi_1$ and $\phi_2$ being the volume fractions of small and large colloids respectively. The structural relaxation time, $\tau_\alpha$, is calculated using $Q_s(t=\tau_\alpha) = 1/e$.

Several studies~\cite{Das2016,Bhowmik2016JSTAT,C7SM01202K} have shown that, $\tau_\alpha$, increases sharply as the fraction, $\rho_{\text{pin}}$, of the pinned particles increases (Fig.~\ref{ratio} (a) and (b)).
At $\rho_{\text{pin}} =0.08$, it is $\sim 300$ times slower compared to the unpinned case at $\phi=0.09$. Therefore, at a fixed $\phi$ increasing $\rho_{\text{pin}}$ is effectively equivalent to compressing the liquid.

\noindent \textbf{Dynamic heterogeneity:} When a glass-forming liquid is compressed, the dynamics become progressively heterogeneous~\cite {Ediger2000,Richert_2002,ThirumalaiRMP2015,Lubchenko07ARPC}.
The degree of dynamic heterogeneity may be characterized by the fluctuations in the self part of the two-point correlation function \cite{Kirkpatrick88PRA,Tonneli05PRE},  
 \begin{equation}
\chi^{s}_{4}(t) = (N-N_p)\left[\left\langle Q_s(t)^2\right\rangle - \left\langle Q_s(t)\right\rangle^2\right],
\label{CHI4}
\end{equation}
where $\left[\left\langle ...\right\rangle\right]$ represent the ensemble average and average over different time origins. 
\begin{figure}[ht!]
\begin{center}
\vskip +0.50cm
\includegraphics[width=0.98\columnwidth]{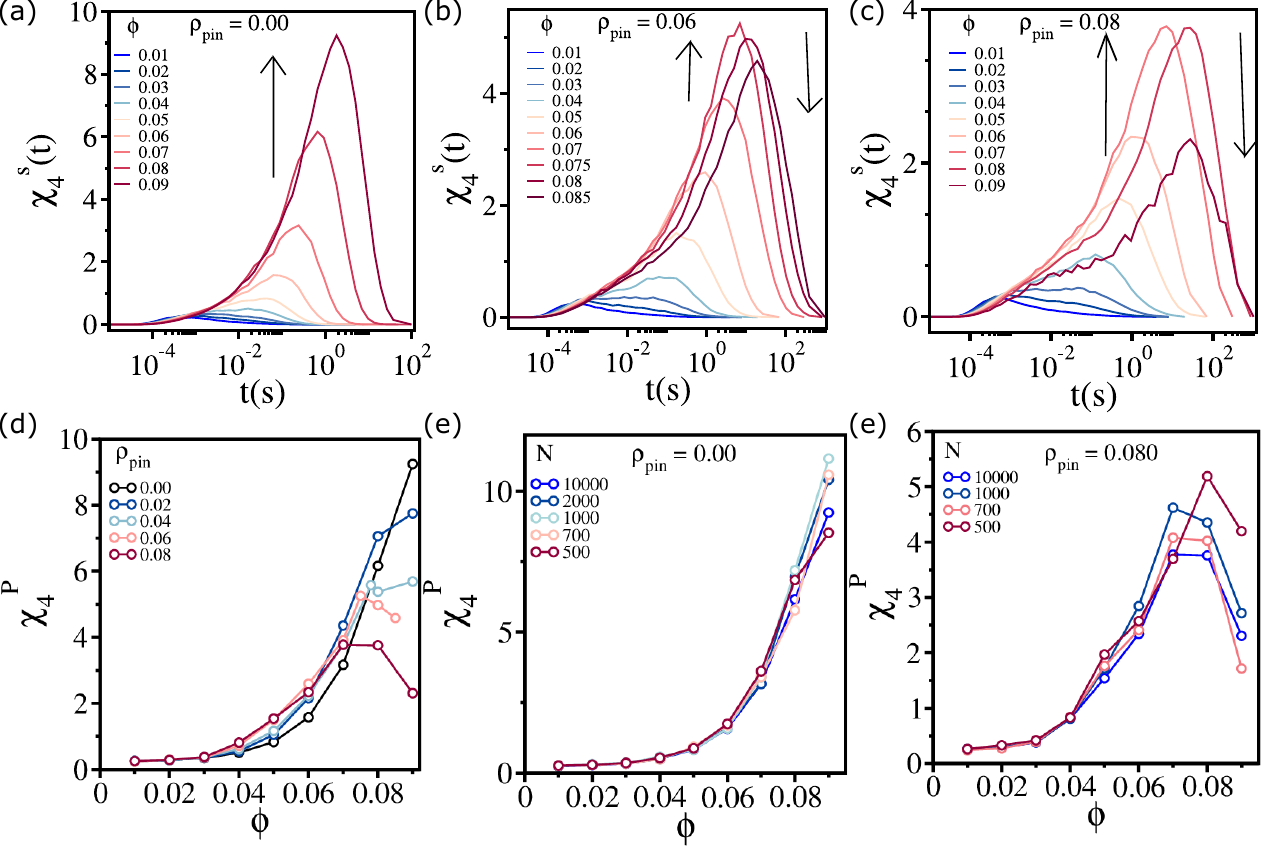}
\caption{ {\bf Non-monotonic growth of $\chi^s_4(t)$: } Four-point dynamic susceptibility $\chi^s_4(t)$ as function of $t$ for $0.01\leq \phi \leq 0.09$ at $\rho_{\text{pin}} = 0.0$ (a),  $\rho_{\text{pin}} = 0.06$ (b) and  $\rho_{\text{pin}} = 0.08$ (c). The arrows in the upward direction indicate the peak in the $\chi^s_4(t)$ increases whereas the downward arrows indicate it decreases. (d) The peak value of $\chi^s_4(t)$ ($\chi^{P}_4$) as a function of $\phi$ for $0.0 \leq \rho_{\text{pin}} \leq 0.08$. (e) $\chi^{P}_4$ as a function of $\phi$ for $500 \leq N \leq 10,000$ with $\rho_{\text{pin}} = 0.0$. (f) Same as (e) except it is for $\rho_{\text{pin}} = 0.08$.}
\label{chi4Peak}
\end{center}
\end{figure}


We find that the four-point function $\chi^s_4(t)$ grows non-monotonically as a function of $t$ when $\phi$ and $\rho_{\text{pin}}$ increase, which indicates that the initial correlation between particles is minimal and grows with time with a maximum at intermediate times. The maximum in $\chi^s_4(t)$, whose amplitude is $\chi_4^P$, appears at $t \simeq\tau_\alpha$. The peak height $\chi^{P}_4$ is directly related to the degree of dynamic heterogeneity~\cite{Bouchaud2005,Biroli2011,berthier2011dynamical}. Fig.~\ref{chi4Peak} (a) shows that at $\rho_{\text{pin}} = 0.0$, $\chi^{P}_4$ grows monotonically as the liquid is compressed toward $\phi_d$ ($\phi_g \approx 0.10$~\cite{Hyun2019}), which is a signature that the dynamics is increasingly heterogeneous as $\phi$ is increased. 


When the number of pinned particles increases,  the maximum in the $\chi^s_4(t)$ 
does not increase monotonically as the $\phi$ increases (Fig.~\ref{chi4Peak} (b) and (c)). For $\rho_{\text{pin}} > 0.02$, $\chi^{P}_4$  as a function of $\phi$  exhibit non-monotonic behavior (Fig.~\ref{chi4Peak} (d)). It increases until $\phi_N$ ($\phi_N$ is the packing fraction where non-monotonicity appears) and then decreases upon further increases in $\phi$. Moreover, for $\phi > \phi_N$, $\chi^{P}_4$ systematically decreases as $\rho_{\text{pin}}$ increases.  Similar observations were made in earlier works~\cite{KobPRE2014,Wei2015JCP,Bhowmik2016JSTAT}. However, the reason for the non-monotonic behavior of $\chi_4^P$ with the increase in $\phi$ is still unclear. 

To assess if the observed behavior is an effect of finite system size, we performed additional simulations by varying N. The qualitative behavior is independent of the system size for unpinned and pinned liquids (Fig.~\ref{chi4Peak} (e) \& (f)). The results shown in Fig.~\ref{chi4Peak} apparently suggest that when the number of pinned particles increases, the dynamically correlated region exhibits non-monotonic behavior as a function of $\phi$, as assessed by $\chi_4^s(t)$.

\begin{figure}[ht!]
\begin{center}
\vskip +0.50cm
\includegraphics[width=0.98\columnwidth]{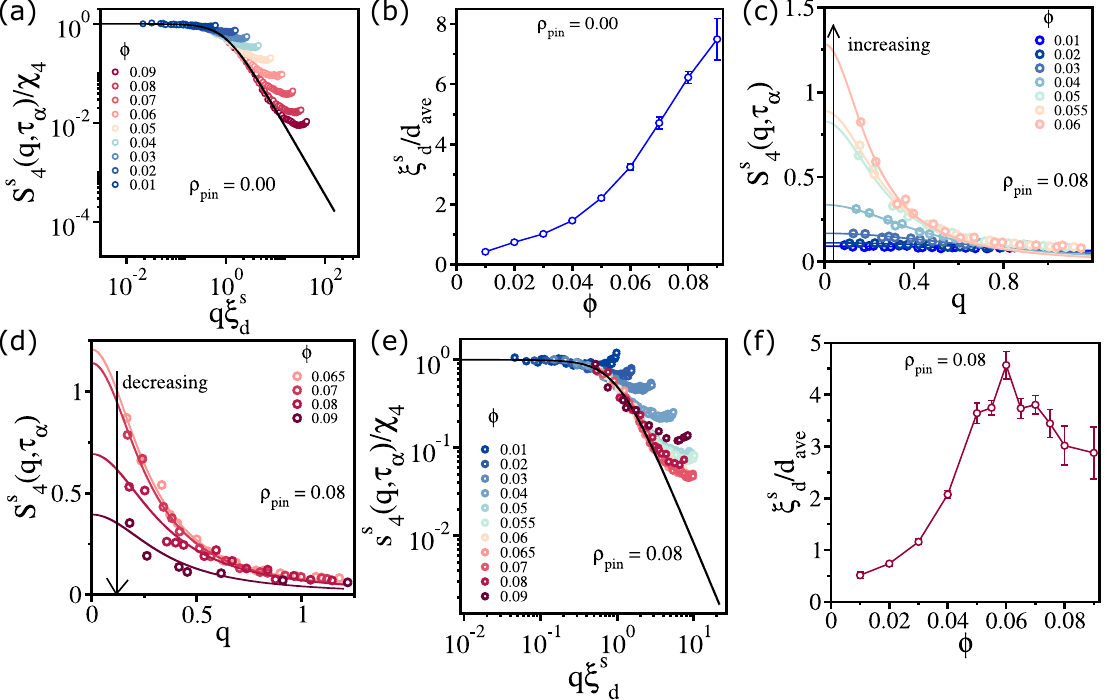}
\caption{ {\bf Non-monotonic growth of the dynamical length from self-correlation function: } (a) $S^s_4(q,\tau_\alpha)$ scaled by $\chi_4$ as a function of $q\xi^s_d$ for $0.01 \leq \phi \leq 0.09$ at $\rho_{\text{pin}} = 0.0$. The solid line is the fit to Eqn.~\eqref{OZ}. (b) $\xi^s_d$ obtained from the OZ fit in (a), as a function of $\phi$. (c)  $S^s_4(q,\tau_\alpha)$ as a function of $q$ for $0.01 \leq \phi \leq 0.06$ at $\rho_{\text{pin}} = 0.08$. (d) Same as (c), except it is for $0.065 \leq \phi \leq 0.09$. (e) Same as (a) but for $\rho_{\text{pin}} = 0.08$. (f) $\xi^s_d$ as a function of $\phi$ for $\rho_{\text{pin}} = 0.08$. }
\label{selfLength}
\end{center}
\end{figure}

\textbf{Dynamic length from single particle dynamics:}
To quantify the length scale associated with the dynamically correlated regions, we first investigated the small wave-number dependence of the self part of the four-point dynamic structure factor $S^s_4(q,t)$~\cite{OZ2}, 
\begin{equation}
S^s_4(q,t) = \left(N-N_p\right)[\langle W^s_0(q,t)W^s_0(-q,t)\rangle -\langle W^s_0(q,t)\rangle^2],
\label{equSelf}
\end{equation} 
where,
\begin{equation}
W^s_0(q,t) = \tfrac{1}{N-N_p}\sum_{ j}^{N-N_p}w_{j}(t)\exp[-i\vec{q}.\vec{r}_j(0)].
\label{w0s}
\end{equation}
In Eq. \ref{w0s}, $w_{j}(t) = \Theta\left(a - |\vec{r}_j(t) - \vec{r}_j(0)|\right)$ with $a = 0.3d_{\text{ave}}$.
The small $q$ dependence of $S^s_4(q,\tau_\alpha)$ is well described by the Ornstein-Zernike (OZ)~\cite{OZ1,OZ2,OZ3,OZ4} equation, 
\begin{equation}
S^s_4(q,\tau_\alpha)\simeq \frac{\chi_4}{1+q^2{\xi^s_d}^2}, 
\label{OZ}
\end{equation} 
where $\xi^s_d$ and $\chi_4$ are fit parameters, $\xi^s_d$ is the size of the dynamically correlated region. The coefficient, $\chi_4$ in Eqn.~\eqref{OZ}, is the value for an infinite system, which is interpreted as the average number of particles belonging to the dynamically correlated region. 

In Fig.~\ref{selfLength} (a), we show the OZ fit to the $S^s_4(q,\tau_\alpha)$ to estimate the dynamic length scale for $\rho_{\text{pin}} = 0.0$. The calculated $\xi^s_d$ grows \textit{monotonically} as a function of $\phi$ (Fig.~\ref{selfLength} (b)), which accords well with FS findings for the harmonic system~\cite{Flenner12NatPhys}. However, for $\rho_{\text{pin}} = 0.08$, $S^s_4(q,\tau_\alpha)$ as a function of $q$ increases with $\phi$ for $\phi \leq 0.06$, and decreases upon further increase in $\phi$ (see Fig.~\ref{selfLength} (c) \& (d)). The calculated  $\xi^s_d$ from the OZ fit (Fig.~\ref{selfLength} (e)) grows \textit{non-monotonically} as a function of $\phi$ (Fig.~\ref{selfLength} (f)). A similar non-monotonic dynamic length scale was reported in previous computational studies~\cite{Kob2012NatPhys,MeiPRE2017,Peng2022PRL} using different interaction potentials as well as in experiments~\cite{Ganapathi18NatComm,HimaNagamanasa15NatPhys}.  
\begin{figure}[ht!]
\begin{center}
\vskip +0.50cm
\includegraphics[width=0.98\columnwidth]{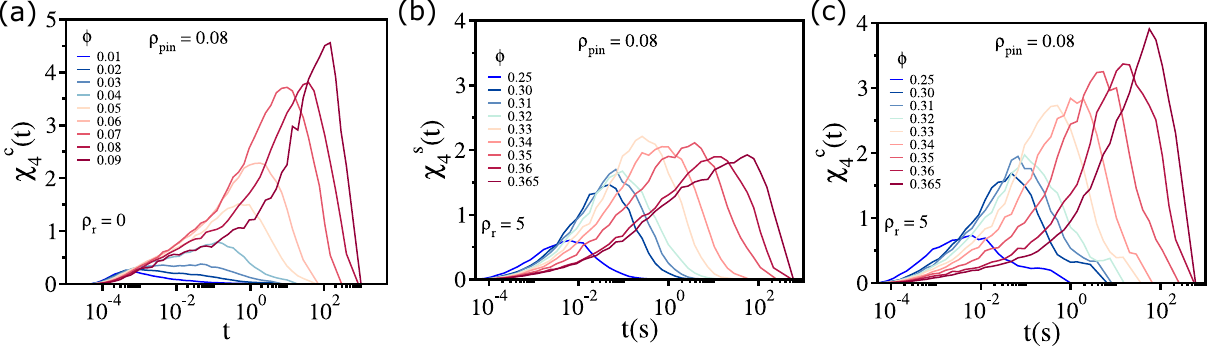}
\caption{ {\bf Monotonic growth of peak in $\chi_4^c$: } \textcolor{black}{(a) The collective four-point dynamic susceptibility $\chi_4^c(t)$ as a function of $t$ for $0.01\leq \phi \leq 0.09$ with $\rho_{\text{pin}} = 0.08$ for $\rho_r = 0$. (b) The self $\chi_4^s(t)$ as a function of $t$ for $0.25\leq \phi \leq 0.365$ with $\rho_{\text{pin}} = 0.08$ for $\rho_r = 5$. (c) Same as (b) but for $\chi_4^c(t)$.}}
\label{collectivechi}
\end{center}
\end{figure}

It is important to emphasize that in the discussion so far, we have calculated the susceptibility ($\chi^s_4(t)$) and the length scale ($\xi^s_d$) from the self part of the correlation function, which is commonly used in the literature~\cite{KobPRE2014,Wei2015JCP,Bhowmik2016JSTAT}. However, the self part of the correlation function does not account directly for how single particle affects the dynamics of other particles over time especially as $\phi$ approaches $\phi_d$. \textcolor{black}{We find that the peak of collective $\chi_4^c(t)$ increases monotonically (see Fig.~\ref{collectivechi}).} 
\begin{figure}[ht!]
\begin{center}
\vskip +0.50cm
\includegraphics[width=0.98\columnwidth]{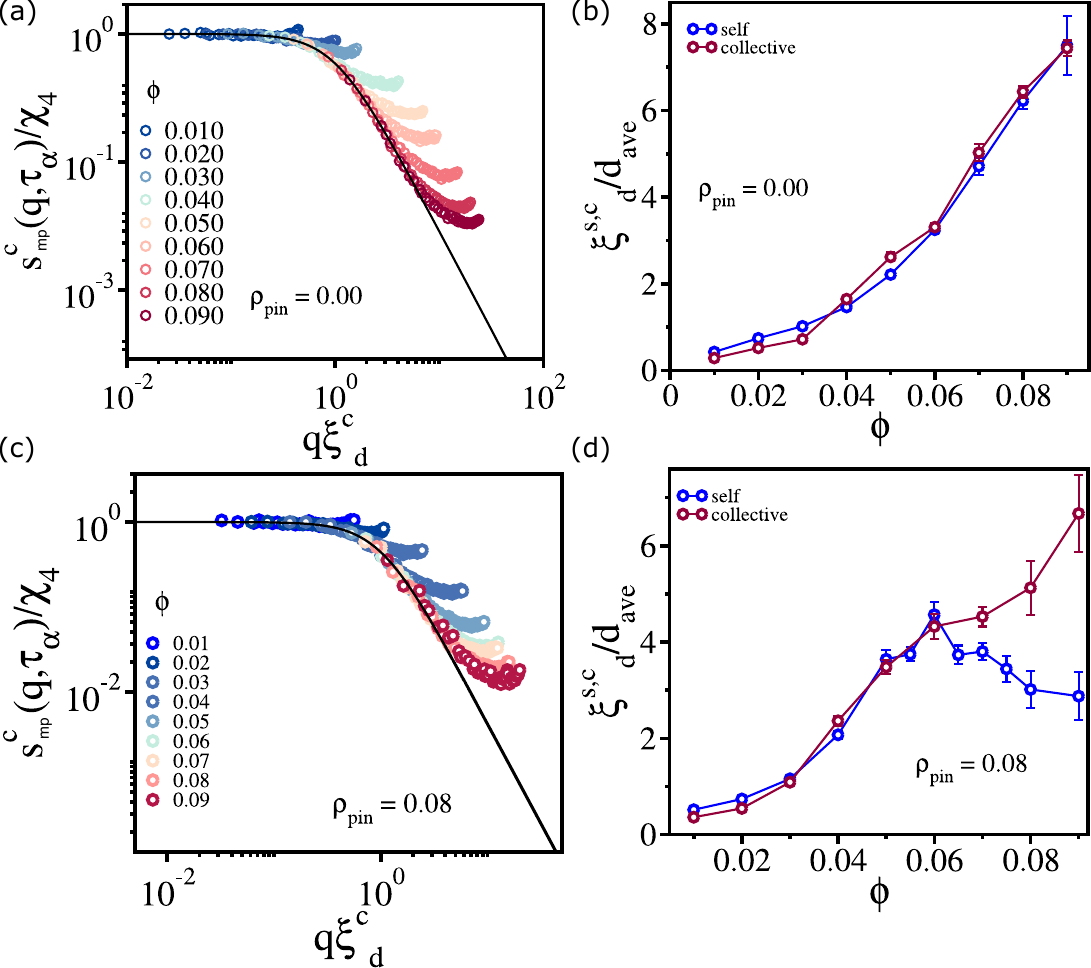}
\caption{ {\bf Monotonic growth of dynamical length from collective correlation function: } (a) $S^c_{mp}(q,\tau_\alpha)$ scaled by $\chi_4$ as a function of $q\xi^c_d$ for $0.01 \leq \phi \leq 0.09$ at $\rho_{\text{pin}} = 0.0$. The solid line is the fit to Eqn.~\eqref{OZ}. (b) $\xi^c_d$ obtained from the fitting in (a) and $\xi^s_d$ obtained from Fig.~\ref{selfLength} (b) as a function of $\phi$ . (c)  Same as (a), but it is for $\rho_{\text{pin}} = 0.08$. (d) Comparison of length scale $\xi^c_d$ obtained from (c) and $\xi^s_d$ from Fig.~\ref{selfLength} (f).}
\label{collectiveLength}
\end{center}
\end{figure}
Therefore, one needs to investigate the collective correlation functions to understand the dynamics of glass-forming liquids correctly~\cite{jack2013dynamical,BiroliJCP,ozawa2015reply}.

\textbf{Dynamic length using multi-particle correlation function:} We calculated the small wave-number dependence of the collective multi-particle dynamic structure factor $S^c_{mp}(q,t)$,
\begin{equation}
    S^c_{mp}(q,t) = \left(N-N_p\right)[\langle W^c_0(q,t)W^c_0(-q,t)\rangle -\langle W^c_0(q,t)\rangle^2],
    \label{equColl}
\end{equation}
where, 
\begin{equation}
W^c_0(q,t) = \tfrac{1}{N-N_p}\sum_{k\neq j}^{N-N_p}w_{kj}(t)\exp[-i\vec{q}.\vec{r}_j(0)].
\label{w0c}
\end{equation}
The two-point overlap function, 
\begin{equation}
    w_{kj}(t) = \Theta\left(a - |\vec{r}_k(t) - \vec{r}_j(0)|\right)
    \label{ProteinOP}
\end{equation}
is an order parameter that was introduced in protein folding to distinguish between folded and unfolded states of globular proteins~\cite{guo1995kinetics}. The OZ fit \textcolor{black}{(see Appendix for details)} to the collective dynamic structure factor $S^c_{mp}(q,t)$ used to calculate $\xi^c_d$  for $\rho_{\text{pin}} = 0.0$ is shown in Fig.~\ref{collectiveLength} (a). The length scale calculated from the $S^c_{mp}(q,t)$ is quantitatively similar to that estimated using $S^s_4(q,t)$ (Fig.~\ref{collectiveLength} (b)). However, for $\rho_{\text{pin}} = 0.08$, $\xi^c_d$ calculated using $S^c_{mp}(q,t)$ grows \textit{monotonically} as a function of $\phi$, whereas the $\xi^s_d$  exhibits a  \textit{non-monotonic} behavior (Fig.~\ref{collectiveLength} (d)). Therefore, we conclude that the self part of the correlation underestimates the overall correlation between particles, resulting in the \textit{non-monotonic} growth of $\xi^s_d$ at non-zero values of $\rho_{\text{pin}}$. 

The maximum in $\xi^s_d$ in Fig.~\ref{selfLength} (f)  is at $\phi \approx \phi_d$, estimated by fitting $\xi^c_d \sim (\phi_d -\phi)^{-\gamma}$ (or to the $\tau_\alpha$ as function of $\phi$ data). \textcolor{black}{We estimate $\phi_d \simeq 0.064$ at $\rho_{\text{pin}}=0.08$, which is less than the value at $\rho_{\text{pin}}=0.0$ ($\phi_d\simeq 0.10$, see Table I in Ref~\cite{Hyun2019}).}
This implies that collective dynamics is negligible at $\phi \lesssim \phi_d$ and become important only when $\phi$ exceeds $\phi_d$ but still less than $\phi_g$. It is worth noting that free energy barrier limited transport starts to be prominent even before $\phi_d$~\cite{Mountain93Physcia}.
\textcolor{black}{We should emphasize that, as highlighted by FS~\cite{Flenner12NatPhys}, the previously utilized generalized point-to-set method captures the nonlinear response of the pinned particles on the dynamics of the unpinned particles, rather than directly measuring the actual multi-particle dynamic correlations. In contrast, the collective $S^c_{mp}(q,t)$ quantifies how slow-moving particles are correlated with one another over a length scale that is \textit{intrinsic} to the liquid.}
\begin{figure}[ht!]
\begin{center}
\includegraphics[width=0.58\columnwidth]{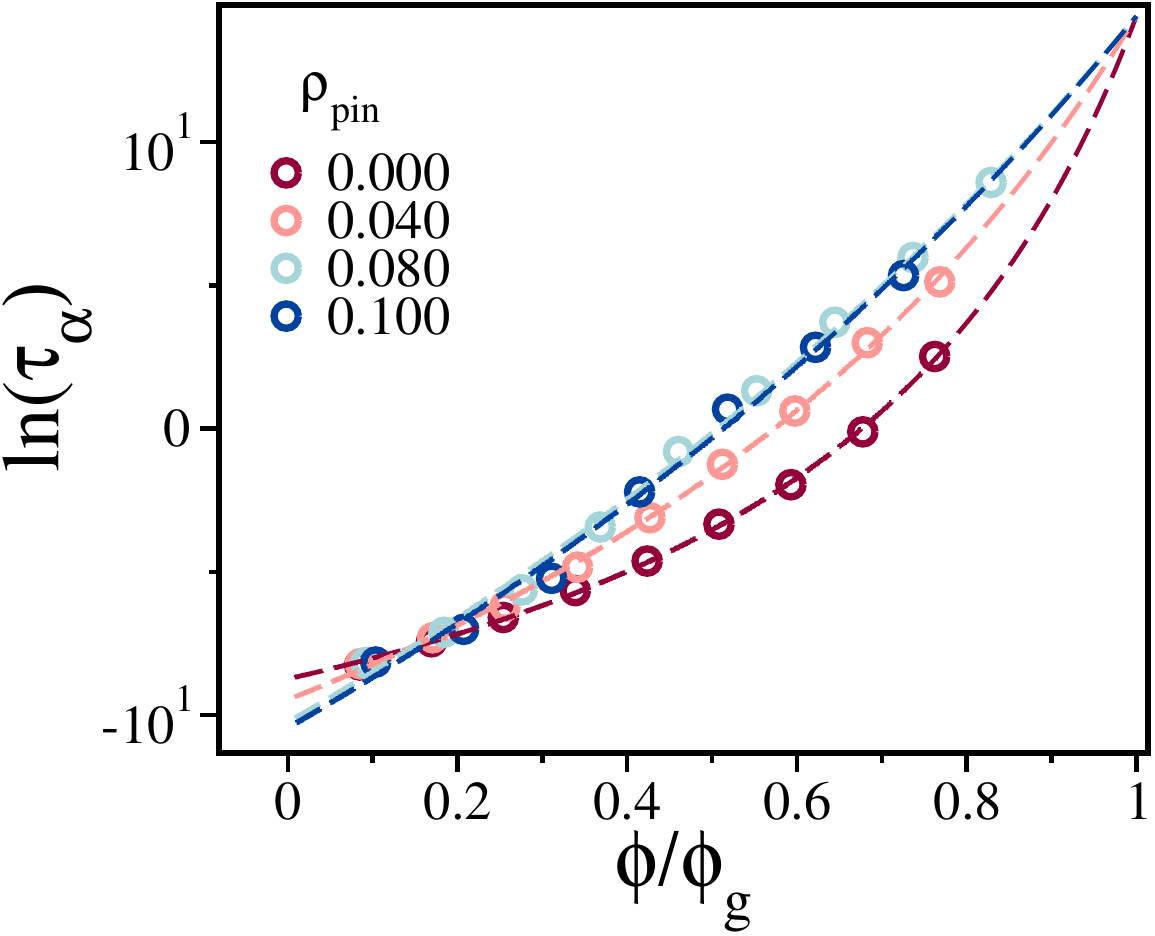}
\caption{ {\bf  Angell plot for various pinning fractions:} \textcolor{black}{Logarithm  of the relaxation time $\tau_\alpha$ as a function $\phi/\phi_g$ for $\rho_r = 0$. Here $N = 1000$.}}
\label{angell}
\end{center}
\end{figure}

\textcolor{black}{It is important to note that the collective length scale $\xi_d^c$ for the unpinned system is larger than $\xi^c_d$ for pinned system (compare Fig.~\ref{collectiveLength} (b) and (d)). This is a result of  pinned particles altering the system drastically-- the pinned system becomes ``strong" liquids (Fig.~\ref{angell}). Therefore, one would expect a slow evolution of the length scale in the pinned system.} 
\begin{figure}[ht!]
\begin{center}
\vskip +0.50cm
\includegraphics[width=0.98\columnwidth]{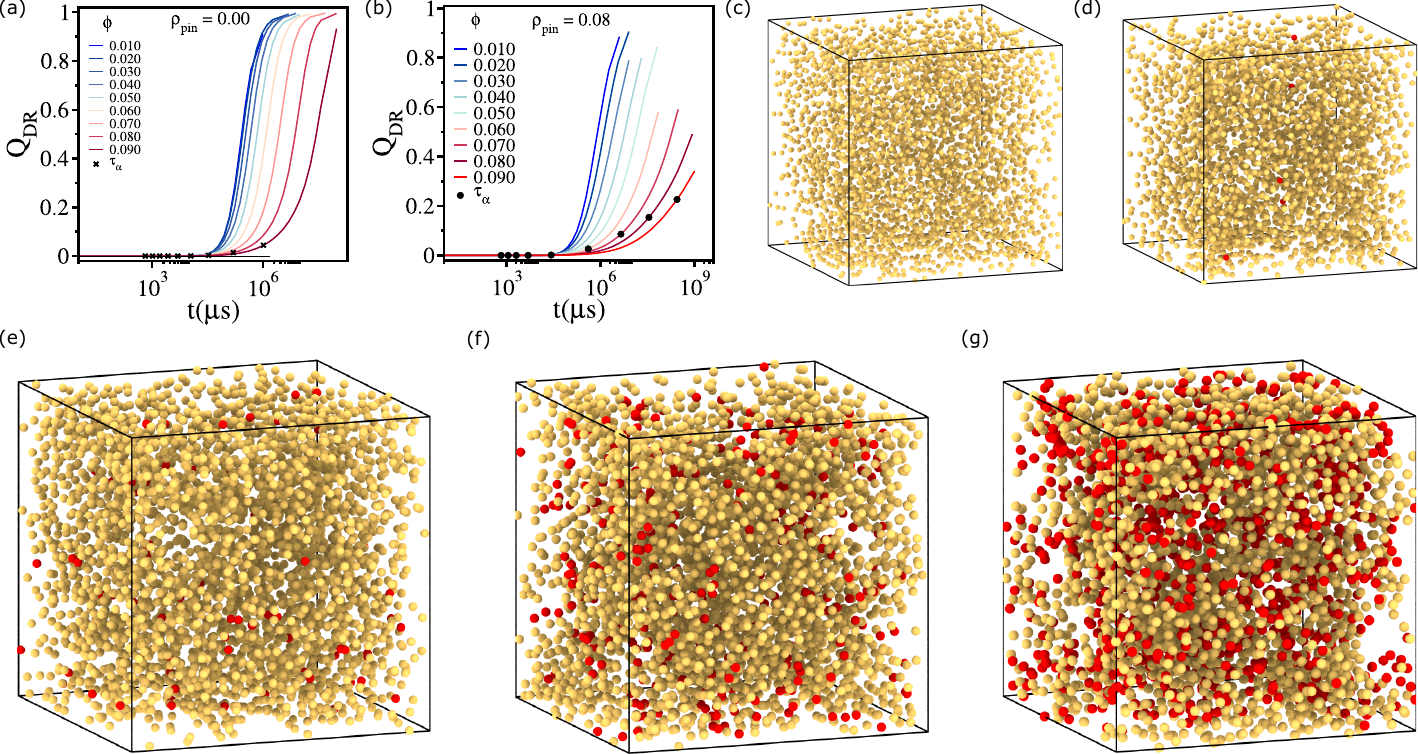}
\caption{ {\bf Self and distinct correlation:} (a) $Q_{DR} = Q_D/Q_c$ as a function of $t$ for $0.01\leq\phi\leq0.09$ with $\rho_{\text{pin}} =0.0$ for $\rho_r=0$. The relaxation time at different $\phi$ is shown as * in the curve. (b) Same as (a) for $\rho_{\text{pin}} =0.08$. Relaxation time at different $\phi$ is shown as the black circle in the curves. Simulation snapshots for self-correlated and distinctly correlated particles at $t = \tau_\alpha$ at $\rho_{\text{pin}} =0.08$ for $\phi=0.03$ (c), $\phi=0.05$ (d), $\phi=0.06$ (e), $\phi=0.07$ (f) and $\phi=0.09$ (g). Yellow particles are self-correlated, and red particles are distinctly correlated.}
\label{distinct}
\end{center}
\end{figure}


To quantify the extent to which $S^s_4(q,t)$ underestimates the overall correlation between particles over time, we first define the collective overlap function $Q_c(t)$,
\begin{align}
    Q_c(t)&=\frac{1}{N-N_p}\left[\left\langle\sum_{i,j=1}^{N-N_p}w_{ij}(t)\right\rangle\right]\nonumber\\
    &= \frac{1}{N-N_p}\left[\left\langle\sum_{i=1}^{N-N_p}w_{i}(t)\right\rangle\right] + \frac{1}{N-N_p}\left[\left\langle\sum_{i\neq j }^{N-N_p}w_{ij}(t)\right\rangle\right] \nonumber \\
    & = Q_{S} + Q_{D}. 
    \label{Qcsplit}
\end{align}
Next, we define $Q_{\text{DR}} = \frac{Q_D}{Q_{c}}$.
For $\rho_{\text{pin}} =0.0$, we find that at $t \simeq\tau_\alpha$, the contribution of $Q_{D}$ to the overall correlation is almost negligible even at $\phi=0.09$ (Fig.~\ref{distinct} (a)). As a consequence, the length scale estimated from the collective correlation function $S^c_{mp}(q,t)$ and the self-correlation function $S^s_4(q,t)$ do not differ (Fig.~\ref{collectiveLength} (b)). This is also reflected in $Q_{DR}$ at $\tau_\alpha$, which increases only by $\sim 5\%$ at $\phi = 0.09$ . In contrast at $\rho_{\text{pin}} =0.08$, $Q_{D}$ is significant ($\gtrsim 20\%$ for $\phi\geq 0.08$) at $t=\tau_\alpha$ when $\phi\geq 0.06$ (Fig.~\ref{distinct} (b)). Strikingly, for $\phi\geq 0.06 \approx \phi_d$, the single particle length $\xi^s_d$ calculated using $S^s_4(q,t)$ starts to decrease upon increasing $\phi$ (Fig.~\ref{selfLength} (f)). 

To illustrate the importance of two-point functions, we define a particle as self-correlated for which $Q_{S} >0.0$ at $t=\tau_\alpha$ and distinctly correlated if $Q_{D} >0.0$ (Eqn.~\eqref{Qcsplit}). As shown in Fig.~\ref{distinct} (c)-(f), a significant number of particles become distinctly correlated when $\phi\geq 0.06$, which is signaled by the increase in $Q_{DR}$. 
We identify this value of $\phi$ as the onset of collective dynamics. 
After the onset of the collective dynamics, one would see an apparent non-monotonic dynamic length scale if calculated from the self part of the correlation function (Fig.~\ref{collectiveLength} (f) and Fig.~\ref{collectiveLengthRho5} (b)). 
\begin{figure}[ht!]
\begin{center}
\includegraphics[width=0.99\columnwidth]{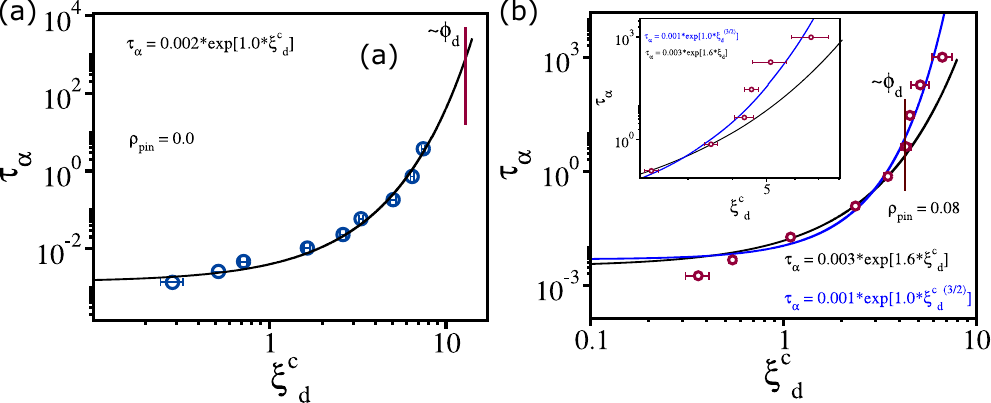}
\caption{ {\bf Dynamical crossover:} (a) Relaxation time $\tau_\alpha$ as a function of dynamic length scales $\xi^c_d$ in a log-log scale for $\rho_r =0$ with $\rho_{\text{pin}} = 0.0$. The solid black line is fit to $\tau_\alpha \sim \exp(k{\xi^c_d})$. The vertical red line indicates the position of $\phi_d$~\cite{Hyun2019}. (b) Same as (a), but for $\rho_{\text{pin}} = 0.08$.  The blue dashed line is fit to $\tau_\alpha \sim \exp(k(\xi^c_d)^{3/2})$ and the black dashed line is fit to $\tau_\alpha \sim \exp(k{\xi^c_d})$. The vertical red line corresponds to the position of $\phi_d$. The inset shows the zoomed-in version for better visualization.} 
\label{crossover}
\end{center}
\end{figure}

\textbf{Dynamic crossover:}
The dependence of $\xi_d^s$ for $\rho_{\text{pin}} = 0.08$ with a maximum at $\phi\sim 0.06$ suggests that there is a change in the dynamics, as noted previously~\cite{Flenner12NatPhys}.   Interestingly, the onset of collective dynamics is also reflected in a log-log plot of $\tau_\alpha$ as a function of $\xi^c_d$ (Fig.~\ref{crossover} (b)). For $\phi \leq 0.06$ (moderately compressed region), the data is well described by $\tau_\alpha \sim \exp[k\xi^c_d]$. In contrast, in the compressed RFOT-dominated regime, the data is better fit by $\tau_\alpha \sim \exp[k(\xi^c_d)^{3/2}]$. We surmise that the apparent non-monotonic behavior in the dynamic length scale is a consequence of a dynamic crossover, where the collective motion of particles starts to dominate, and activated processes control transport~\cite{KirkpatrickPRA1989}. 
In unpinned systems, there is no discernible crossover in the dynamics in the studied range of $\phi$ (Fig.~\ref{crossover} (a)). This is because for an unpinned system,  the range of values of $\phi$'s that we can reliably simulate is not sufficient enough to explore RFOT behavior. 
We expect that for the systems in the absence of pinned particles, one might also see such dynamic crossover if the packing fraction is increased, as illustrated in Fig.~\ref{distinct} (a). \textcolor{black}{ The biggest and the most significant advantage of the pinning protocol, preferably by self-pinning, is that the crossover to the RFOT-dominated regime may be simulated.} 
 
\begin{figure}[ht!]
\begin{center}
\includegraphics[width=0.96\columnwidth]{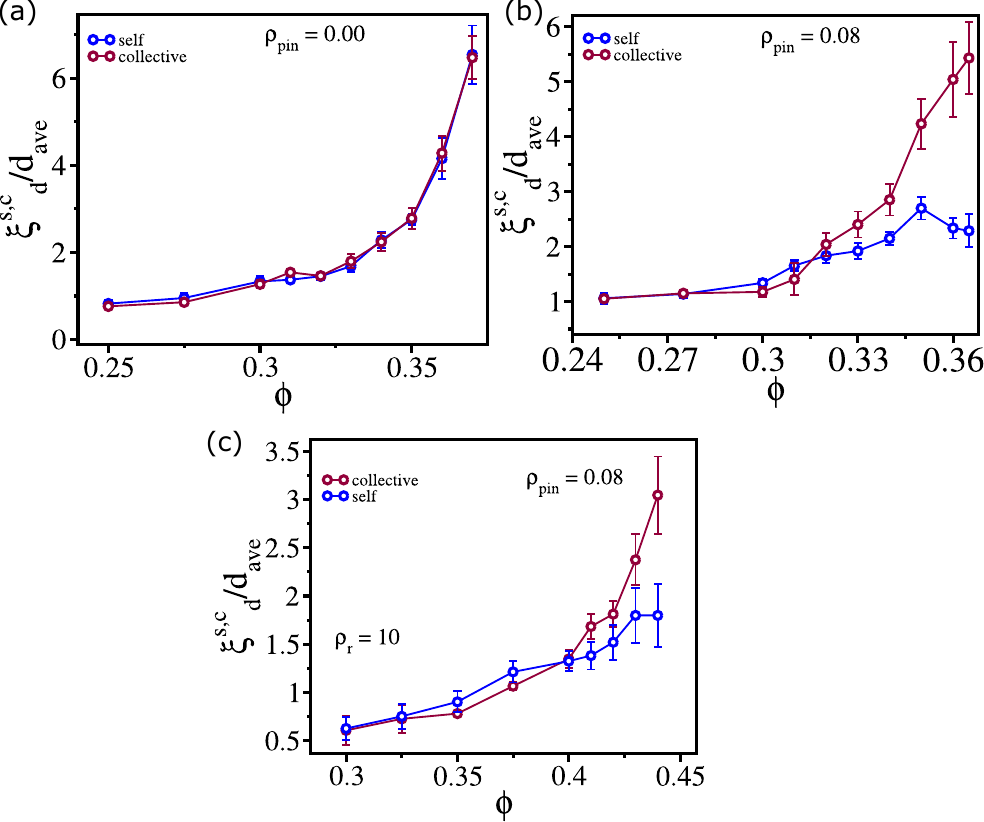}
\caption{ {\bf Comparison of dynamic length  from collective and self-correlation function for $\rho_r = 5$}: Plots of $\xi^s_d$ and $\xi^c_d$ as a function of $\phi$ for $\rho_r =5$ using $S_4^s(q,t)$ and $S^c_{mp}(q,t)$ respectively with $\rho_{\text{pin}} = 0.00$ (a) and with $\rho_{\text{pin}} = 0.08$ (b). \textcolor{black}{(c) Same as (b) but for $\rho_r = 10$}.} 
\label{collectiveLengthRho5}
\end{center}
\end{figure}
 \textbf{Salt effects on the dynamic length:}
The interaction potential (Eqn. \eqref{DLVO}) between charged colloidal particles may be altered by adding monovalent salts. Because salt effects can be probed experimentally, it is of interest to predict the changes in the dynamics at $\rho_r \ne 0$.  We performed simulations with $\rho_r =5$ and $10$ (Eqn.~\eqref{rhor}) to assess the extent to which $\xi^c_d$ and $\xi^s_d$ change as a function of $\phi$. As before, we fit  Eqn.~\eqref{equColl} and Eqn.~\eqref{equSelf} to the OZ equation.  
\textcolor{black}{For $\rho_r =5$ and $\rho_{\text{pin}}=0.0,$} the $\xi_d^s$ and $\xi_d^c$ grows \textit{monotonically} as a function of $\phi$ (Fig.~\ref{collectiveLengthRho5} (a)). For $\rho_{\text{pin}} = 0.08$,  $\xi_d^c$ increases \textit{monotonically} but $\xi^s_d$ grows \textit{non-monotonically} as a function of $\phi$ (Fig.~\ref{collectiveLengthRho5} (b)).  It is worth noting that the peak in $\xi_d^s$ is at $\phi \approx 0.35$, which is much higher than the value at $\rho_r =0$ because $\phi_d$ depends on the salt concentration (see Table 1 in \cite{Hyun2019}). \textcolor{black}{It is worth noting that peak in $\xi_d^s$ for $\rho_r =0$ is much more prominent than for $\rho_r =5$ (compare Fig.~\ref{collectiveLength} (d) and Fig.~\ref{collectiveLengthRho5} (b)). For $\rho_r =10 $, there is a sign of non-monotonicity in $\xi_d^s$  at $\phi\sim 0.44$ (Fig.~\ref{collectiveLengthRho5} (c)) although it is not as clear as $\rho_r =0$ or $\rho_r =5$. This is because as the salt concentration increases the effective range of interaction potential decreases. Therefore, the effect of pinned particles on the dynamics of unpinned particles will be weaker at high salt concentrations. Thus, one needs to simulate high $\rho_{\text{pin}}$ or high packing fractions to see similar prominent peak in $\xi^s_d$ as $\rho_r =0$. } 

\section{Conclusions}
We performed Brownian dynamic simulations of a mixture of charged micron-sized colloidal particles to create a consistent method to extract dynamic length in glass forming systems with and without random pinning.  Our primary finding is that the collective dynamic length ($\xi_d^c$), calculated using the small $q$ limit of the multi-point structure factor $S^c_{mp}(q,t)$ (Eqn.~\eqref{equColl}), which is well fit to the OZ function \textcolor{black}{at small $q$}, increases monotonically in both the pinned and unpinned systems.  Given this finding, it is worth pondering about the utility of the RP method, especially when a large perturbation like an amorphous wall \cite{Kob2012NatPhys,HimaNagamanasa15NatPhys,HockyPRE2014} is used to estimate the dynamic length.  Even though the RP method, with different ways of implementing pinning, has provided insights into the nature of the ideal glass transition, the undesired consequences (unknown non-nonlinear response of unpinned particles on the mobile particles and breakage of translational invariance) call into question its ultimate utility. \textcolor{black}{In binary mixture of charged colloidal particles the characteristics of the system change drastically due to the presence of externally imposed disorders. Clearly, the fragility of the system changes by a factor of $\sim 4-5$ for $8\%$ pinning (see Fig.~\ref{angell}). Therefore, the length scale obtained by pinning particles is not the length scale associated with the actual liquid under consideration.} As shown here and established previously by FS~\cite{Flenner12NatPhys}, it suffices to extract the dynamic length using Eqn. \eqref{equColl} or Eqn. \eqref{OZ} in the absence of random pinning. \textcolor{black}{We hasten to add that hints in the crossover behavior in the relaxation times to the RFOT-dominated regime can most readily be revealed by probing the dynamics of the constrained liquids by random pinning methods \cite{Cammarota8850}.  }


Because the non-monotonic dependence of $\xi_d^s$ appears when collective dynamics start to become significant, we surmise that the self-four-point dynamic structure factor $S^s_4(q,t)$ underestimates the overall correlation between particles. In contrast, $\xi_d^c$ calculated using multi-particle structure factor $S^c_{mp}(q,t)$ increases \textit{monotonically} as a function of $\phi$. 
It is important to note that, although pinning does decrease the magnitude of $\xi_d^s$ (compare panels (b) and (f) in Fig. \ref{selfLength}), the qualitative behavior is unaltered at $\phi < \phi_d$.   
Collective dynamics, which is accompanied by the change in the shape (fractal to compact) of the cooperative regions~\cite{Stevenson06NatPhys}, is prominent only at volume fraction that exceeds  $\phi_d$. The crossover in the dynamics, where collective motions of particles become important, is also reflected not only in that shape of the cooperative regions but also in the dependence of the $\tau_\alpha$ on the dynamic length.  The RFOT-dominated regime ($\tau_\alpha \sim \exp[k(\xi_d^c)^{3/2}]$) appears roughly at the packing fraction where collective dynamics start to emerge. 

The prediction that the collective dynamic length may be estimated in the presence or absence of random pinning can be verified in experiments. All it requires is the reanalysis of the video microscopy data reported previously~\cite{Ganapathi18NatComm} using the method introduced here. Rather than measure single-particle dynamics, it would be necessary to probe correlated dynamics using the order parameter proposed here.  



\textbf{Acknowledgements:} DT is grateful to Peter Rossky for discussions on the history of MD simulations of proteins. We are grateful to Hyun W. Cho for help in the early stages of this work. This work was supported by a grant from the National Science Foundation (CHE 2320256) and the Welch Foundation through the Collie-Welch chair (F0019). 

\clearpage
\appendix
\setcounter{secnumdepth}{0}
\section{Appendix}
\noindent \section{Calculation of a growing length using $S^c_{mp}(q,t)$}
\begin{figure}[ht!]
\begin{center}
\includegraphics[width=0.98\columnwidth]{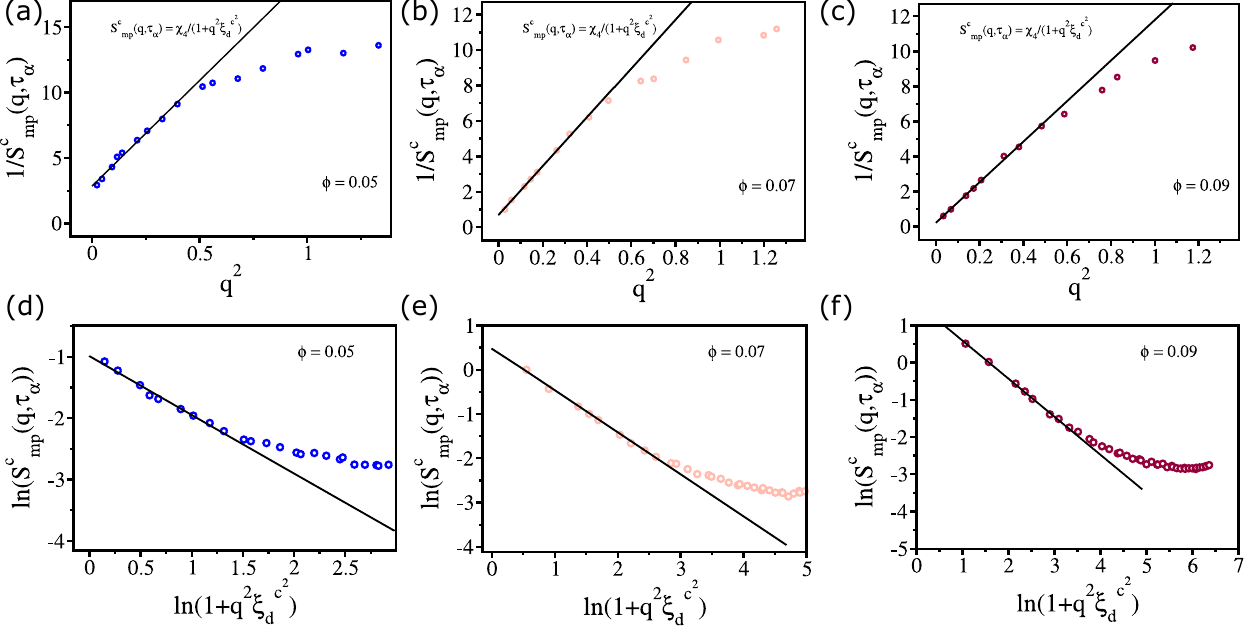}
\caption{ {\bf  Estimation of length scale from $S_4(q,t)$:} \textcolor{black}{(a)-(c) $1/S^c_{mp}(q,\tau_\alpha)$ as a function of $q^2$ for $\phi =0.05, 0.07$ and $0.09$ respectively. The solid lines are linear fit to the function $1/S^c_{mp}(q,\tau_\alpha) = 1/\chi_4 + q^2{\xi^c_d}^2/\chi_4$. (d)-(e) $\ln( S^c_{mp}(q,\tau_\alpha))$ as a function of $\ln (1+q^2{\xi^c_d}^2)$ for $\phi =0.05, 0.07$ and $0.09$ respectively. The solid lines are linear fit. The results are for $\rho_r = 0$ with $\rho_{\text{pin}} =0.00$.}}
\label{s4qt0}
\end{center}
\end{figure}

\begin{figure}[ht!]
\begin{center}
\includegraphics[width=0.98\columnwidth]{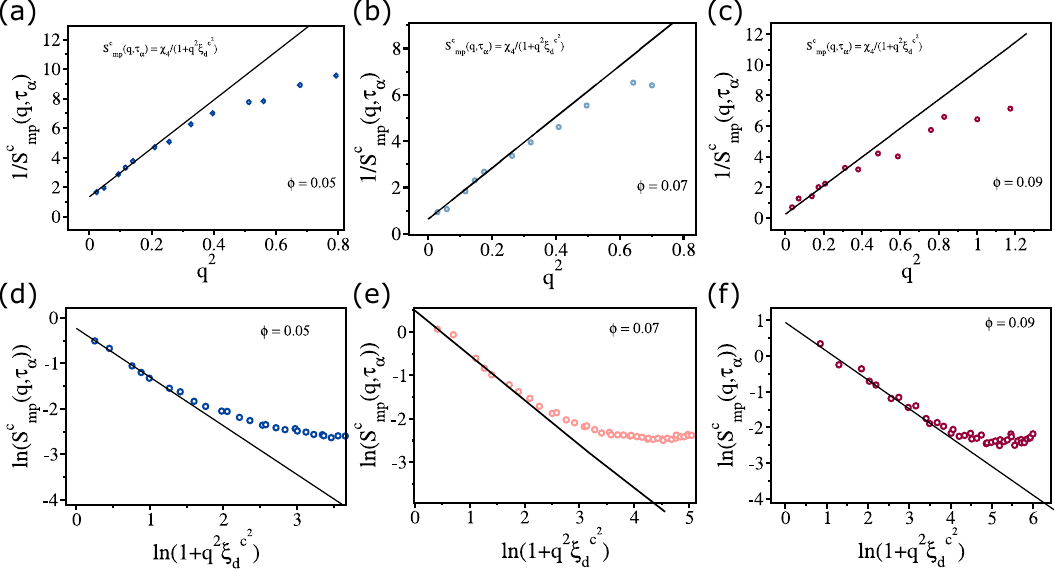}
\caption{ {\bf  Estimation of length scale from $S_4(q,t)$:} \textcolor{black}{(a)-(c) $1/S^c_{mp}(q,\tau_\alpha)$ as a function of $q^2$ for $\phi =0.05, 0.07$ and $0.09$ respectively. The solid lines are linear fit to the function $1/S^c_{mp}(q,\tau_\alpha) = 1/\chi_4 + q^2{\xi^c_d}^2/\chi_4$. (d)-(e) $\ln( S^c_{mp}(q,\tau_\alpha))$ as a function of $\ln (1+q^2{\xi^c_d}^2)$ for $\phi =0.05, 0.07$ and $0.09$ respectively. The solid lines are linear fit. (a)-(f) are for $\rho_r =0$ with $\rho_{\text{pin}} =0.08$.}}
\label{s4qt}
\end{center}
\end{figure}

\textcolor{black}{To demonstrate that  $\xi^c_d$ can be reliably calculated  from the low-$q$ values of $S^c_{mp}(q,\tau_\alpha)$, we employed two methods. As illustrated in Fig.~\ref{s4qt0} (a)-(c) (for $\rho_{\text{pin}}=0.0$) and Fig.~\ref{s4qt} (a)-(c)(for $\rho_{\text{pin}}=0.08$) , the inverse function $1/S^c_{mp}(q,\tau_\alpha)$ plotted as a function of $q^2$ is linear at low $q$ values, which shows  that extrapolating $S^c_{mp}(q,\tau_\alpha)$ to the $q \to 0$ limit is robust. Specifically, at low $q$ values, at least 6–7 data points were well-fit to a linear function, allowing for accurate calculation of the length scale.}

\textcolor{black}{To further validate this approach, we plotted $\ln(S^c_{mp}(q,\tau_\alpha))$ as a function of $\ln(1 + q^2{\xi^c_d}^2)$ using the estimated values of $\xi^c_d$. The low-$q$ data points also showed a strong linear fit to this function, confirming the reliability of the estimation (Fig.~\ref{s4qt0} (d)-(f) and Fig.~\ref{s4qt} (d)-(f)). The results shown in Fig.~\ref{s4qt} are for $\rho_{\text{pin}} = 0.0$ and Fig.~\ref{s4qt} are for $\rho_{\text{pin}} = 0.08$.}

\bibliographystyle{unsrt}
\bibliography{Results}
\end{document}